\pdfoutput=1
\documentclass[aps, prb, 10pt, superscriptaddress,twocolumn]{revtex4-1}

\usepackage{graphicx}
\usepackage{amsfonts}
\usepackage{amsmath}
\usepackage{hyperref}

\usepackage{amssymb}

\usepackage{comment}

\def\Tr{\mbox{Tr}\,}
\newcommand{\la}{\label}
\newcommand{\be}{\begin{equation}}
\newcommand{\ee}{\end{equation}}
\newcommand{\bea}{\begin{eqnarray}}
\newcommand{\eea}{\end{eqnarray}}
\newcommand{\p}{\partial}

\newcommand{\cA} {\mathcal{A}}

\newcommand{\cD} {\mathcal{D}}

\newcommand{\cL} {\mathcal{L}}

\newcommand{\cO} {\mathcal{O}}

\newcommand{\cS} {\mathcal{S}}

\newcommand{\bx} {\mathbf{x}}

\newcommand{\sign}{\mathrm{sgn}} 
\newcommand{\mq}{\mathrm{q}}

\newcommand\x{\mathbf{x}}
\newcommand\q{\mathbf{q}}
\newcommand\+{\dagger}
\renewcommand\d{\partial}

\usepackage{color}

\begin{document}


\title{Fractional quantum Hall systems near nematicity: bimetric theory,\\  composite fermions, 
and Dirac brackets}

\author{Dung Xuan Nguyen}
\affiliation{Rudolf Peierls Center, Oxford University, Oxford OX1 3NP, United Kingdom}

\author{Andrey Gromov}
\author{Dam Thanh Son}
\affiliation{Kadanoff Center for Theoretical Physics, University of Chicago, Illinois 60637, USA}

\begin{abstract}
We perform a detailed comparison of the Dirac composite fermion and the recently proposed bimetric theory for a quantum Hall Jain states near half filling. By tuning the composite Fermi liquid to the vicinity of a nematic phase transition, we find that the two theories are  equivalent to each other. 
We verify that the single mode approximation for the response functions and the static structure factor becomes reliable near the phase transition.  We show that the dispersion relation of the nematic mode near the phase transition can be obtained from the Dirac brackets between the components of the nematic order parameter.  The dispersion is quadratic at low momenta and has a magnetoroton minimum at a finite momentum, which is not related to any nearby inhomogeneous phase.

\end{abstract}

\maketitle

\section{Introduction}

Fractional quantum Hall (FQH) effect is notorious for defying analytical treatment due to the strongly interacting nature of the phenomenon. Various approaches have been proposed for a quantitative treatment of the FQH problem. The first successful approach is based on trial wavefunctions~\cite{1983-Laughlin, haldane1985periodic, moore1991nonabelions, Read1999} and gives many insights into exotic properties of the FQH effect. The approach however, provides little information about excited states.
In a seminal paper\cite{1986-GirvinMacDonaldPlatzman}, Girvin, MacDonald and Platzman tried to address this shortcoming by constructing a wavefunction of an excited density wave state, the Girvin--MacDonald--Platzman (GMP) mode.  Assuming a single mode approximation (SMA), it is possible to estimate the value of various gaps and calculate the observables such as the projected static structure factor, the shear modulus \cite{chui1986shear} and the optical absorption spectrum\cite{1986-GirvinMacDonaldPlatzman}. However, in contrast to the trial ground state, which is an exact eigenstate of a model Hamiltonian, there is no Hamiltonian for which the GMP construction gives an exact energy eigenstate.

Recently the GMP mode has attracted renewed attention. Haldane\cite{haldane2009hall} suggested that the long-wavelength part of the mode is a degree of freedom of geometric nature that has been missed in previous treatments of the FQH effect. Maciejko et al.\cite{maciejko2013field} and You et al.\cite{you2014theory,you2016nematic} considered a nematic phase transition\cite{Xia2011evidence,regnault2016evidence,samkharadze2016observation} where the GMP mode plays the role of the order parameter and the SMA can be reliable.  The low-energy effective theories that these authors constructed disagree quantitatively from each other. The discrepancy was resolved in a ``bimetric'' theory of the FQH states, proposed recently by two of us~\cite{gromov2017bimetric}. The spirit of this theory is quite close to Haldane's original suggestion: the only degree of freedom is the massive spin-$2$ excitation, corresponding to the long-wavelength part of the GMP mode, with a nonlinear, covariant action.
 The theory successfully reproduces the symmetries and correlation functions known from first quantized approaches\cite{CLW,can2014geometry,Gromov-galilean}, and is consistent with constraints from particle-hole symmetry.

One objective of the present paper is to bring the bimetric theory into contact with the idea of the composite fermion\cite{jain1989composite}, a very successful approach to FQHE  \cite{LopezFradkin,halperin1993theory} which provides a natural explanation for the Jain's sequences and the Fermi-liquid behavior of half-filled Landau level. The theory has been recently revised to satisfy particle-hole symmetry at half filling. The resulting Dirac composite fermion (DCF) theory\cite{Son2015composite} has recently been analyzed and found to provide results consistent with particle-hole symmetry and other constraints, but different from the old Halperin--Lee--Read (HLR) theory~\cite{NGRS} (see however Ref.~[\onlinecite{Wang:2017cmz}]). From the point of view of the composite fermion theory, the nematic phase transition occurs due to an instability of the Fermi liquid when the Landau parameter $F_2$ approaches the limit of stability, $-1$.  Then, the quadrupole deformation of the composite Fermi surface becomes soft, and plays the role of the nematic order parameter. In Ref.~[\onlinecite{NGRS}] a bosonized approach to the composite Fermi liquid theory has been developed, in which the Landau parameters can be set to arbitrary values, in particular, those corresponding to the vicinity of a nematic phase transition. 

Our analysis shows, in particular, that the bimetric theory and the DCF theory agree with each other wherever they can be compared. The bimetric theory and the DCF theory have different and complementary regimes of validity.  The bimetric theory can be formulated for any value of the filling factor (including, e.g., $\nu=1/3$) and is applicable near the phase transition and for long wavelengths. The DCF approach works only for filling factors near $\nu=1/2$; in this paper, we consider the two Jain sequences, which we term the ``negative'' and ``positive'' sequences,
\begin{equation}
\nu = \nu_-^N \equiv \frac N{2N+1} ~ \textrm{and} ~  \nu= \nu_+^N = \frac {N+1}{2N+1}\,,
\end{equation} 
  at large $N$.
The bimetric theory and the DCF theory can be compared at large $N$ and small $qN$.  At large $N$, one can obtain additional information from the DCF theory in the momentum regime $Nq\ell\sim1$, where the momentum expansion has already broken down.

The structure of this paper is as follows. In Sec.~\ref{sec:algebraic} we provide an overview of the correspondence between the bimetric theory and the DCF theory. We also describe an approach, based on the Dirac brackets, to the spectrum of the lowest excitation near the nematic phase transition.
In Sec.~\ref{sec:bimetric} we briefly review the non-linear bimetric theory with attention to the Jain sequences.
We switch to the DCF theory in Sec.~\ref{sec:DCFL}. After providing a brief introduction, we derive the quadratic effective action for the dynamics of the quadrupolar (spin-$2$) deformation of the Fermi surface. We find the complete agreement between the quadratic effective action for the bimetric theory and the DCF theory near the nematic phase transition. In Sec.~\ref{sec:linres_DCF}, we calculate the linear response functions at finite frequency and the dispersion relation of the spin-$2$ mode. These calculations can be done in either theory, since the two theories agree upon linearization.
Finally, in Sec.~\ref{sec:spin3}, we verify the validity of the SMA for the static structure factor at the phase transition.
Section \ref{sec:concl} contains concluding remarks.

\section{Algebraic considerations}
\label{sec:algebraic}

A fair amount of information about the nematic phase transition can be obtained without going into field-theoretical details of the bimetric and DCF theories.  This section emphasizes the algebraic, kinematic aspects of the problem. We first map the degrees of freedom between the two theories.  We establish the connection between the Berry phase terms in the bimetric Lagrangian with the commutators and the Dirac brackets on the DCF side.  The computation of the Dirac brackets allows us to determine the nontrivial shape of the dispersion curve near the phase transition.

\subsection{Mapping between the bimetric and DCF theories}

In bimetric theory, the low-energy degree of freedom (the nematic order parameter) is a dynamic unimodular metric $\hat g_{ij}(x)$, which can be parametrized by the real and imaginary parts of a complex field $Q=Q_1+iQ_2$
\be\la{hatgQ}
\hat g_{ij} = \exp \begin{pmatrix}
   Q_2   & Q_1 \\
   Q_1       & -Q_2
\end{pmatrix}\,,\quad Q = Q_1+ iQ_2\,.
\ee

In the DCF theory, the low-energy degree of freedom is the shape of the Fermi surface.  At each spacetime point $x$, the Fermi surface is a closed curve in the two-dimensional momentum space. Using polar coordinates $(k,\theta)$, such a curve can be parametrized by specifying the dependence of the radial coordinate over the angular coordinate; $k=k_F(\theta)$.  For small perturbations, one expands
\begin{equation}\label{eq:FS-shape}
k_F(\theta)=k_F+\!\!\sum_{n=-\infty}^{\infty} \! u_ne^{in\theta}.
\end{equation}
Near the nematic phase transition, quadrupole deformation of the Fermi surface $u_{\pm2}$, which corresponds to a Fermi surface of elliptical shape, becomes the lightest degree of freedom. 
We now identify\cite{you2014theory} this quadrupolar deformation of the Fermi surface with the dynamic metric $\hat g_{ij}$ in the bimetric theory.  More precisely, at linear order, we identify $u_{\pm2}$ with $Q$ and $\bar Q$.  To make the correspondence precise, one turns on a small perturbation of the external metric $g_{ij}=\delta_{ij}+h_{ij}$, $h_{ij}\ll1$, and compares the ground state (the state
with lowest energy) in the two descriptions.  In bimetric theory, the potential energy term favors $\hat g_{ij}=g_{ij}$, i.e.,  $Q_1=h_{12}$ and $Q_2=h_{11}$.  In the DCF theory, the ground state is a state where
the Fermi surface is deformed to
\begin{equation}
  g^{ij} k_i k_j = k_F^2.
\end{equation}
For small metric perturbations, $g_{ij}=\delta_{ij}+ h_{ij}$, the shape of the Fermi surface is then
\begin{equation}
  k_F(\theta) = k_F \left( 1+ \frac{h_{11}}2\cos2\theta + \frac{h_{12}}2\sin2\theta\right),
\end{equation} 
which corresponds to 
\begin{equation}
  u_2 = \frac{k_F}4 (h_{11}-i h_{12}), \qquad u_{-2} = \frac{k_F}4 (h_{11}+ih_{12}).
\end{equation}
Thus, one obtains the following relationship between the two theories
\begin{equation}\label{eq:mapping-firstpass}
   u_2 = -\frac i4 k_F Q, \qquad u_{-2} = \frac i4 k_F \bar Q .
\end{equation}

\subsection{Commutator between components of nematic order parameter}

One of the most important features of bimetric theory is the presence of a term with one time derivative in the action, $\bar Q\dot Q$, indicating that $Q$ and $\bar Q$ do not commute.  The coefficient of this ``Berry phase'' term has been a point of controversy: the values suggested in Refs.~[\onlinecite{maciejko2013field}] and [\onlinecite{you2014theory}] do not agree.  In Ref.~[\onlinecite{gromov2017bimetric}], this coefficient was determined, in particular, by requiring that the nematic mode is solely responsible for the difference between the values of the Hall viscosity at low and high frequencies. The result is 
\begin{equation}
  [\bar Q(\x), \,  Q(\x')] = \frac{16\pi}{\nu(\mathcal S-1)} \ell^2\delta(\x-\x') + \cdots
\end{equation}
where $\ell$ is the magnetic length: $\ell=1/\sqrt B$ where $B$ is the external magnetic field, $\mathcal S$ is the shift of the quantum Hall state, and $\cdots$ are higher spatial derivatives of $\delta(\x-\x')$, to which we will return later. For the Jain state with $\nu=N/(2N{+}1)$, $\mathcal S=N+2$, and up to the next-to-leading order in $N$, one has
\begin{equation}\label{eq:QQbar-comm}
  [\bar Q(\x), \, Q(\x')] =  \frac{64\pi}{2N{+}1} \ell^2\delta(\x-\x').
\end{equation}

On the other hand, in the DCF liquid, the shape fluctuations of the Fermi surface satisfy the commutation relation\cite{Haldane:1994,golkar2016higher}
\begin{multline}\label{eq:commu}
[u_n(\mathbf{x}),u_{n'}(\mathbf{x}')]= -\frac{2\pi}{k_F^2} \big(n b\delta_{n+n',0}\\
	+ ik_F\delta_{n+n',-1}\partial_{\bar{z}}  + ik_F\delta_{n+n',1}\partial_{z}\big)\delta(\mathbf{x}-\mathbf{x}')\,.
\end{multline}
where $b$ is the magnetic field acting on the composite fermion, and $\d_z$ and $\d_{\bar z}$ are defined in Appendix \ref{sec:conventions}.
The commutator between the quadrupole modes is thus
\begin{equation}\label{eq:u2-comm}
  [u_{-2}(\x), \, u_{2}(\x')] = \frac{4\pi b}{k_F^2} \delta(\x-\x'). 
\end{equation}
With the mapping (\ref{eq:mapping-firstpass}), Eqs.~(\ref{eq:QQbar-comm}) and (\ref{eq:u2-comm}) agree with each other, if
\begin{equation}
  k_F^2 = 2\sqrt{Bb\nu(\mathcal S-1)}\,.
\end{equation}
Substituting the values for the Jain's state under consideration, we see that $k_F=\sqrt{B}$ up to $O(1/N^2)$.   The corresponding density of the composite fermions is $B/4\pi$,  consistent with the Dirac composite fermion theory, but deviates from its value in the HLR theory by $O(1/N)$.  The nematic phase transition is another circumstance where the predictions of the Dirac composite fermion theory and the HLR theory (in the versions used so far) differ from each other to $1/N$ order.

\subsection{$q^2$ correction to the commutator, Dirac brackets, and the shape of the dispersion curve}

It has been argued further in Ref.~[\onlinecite{gromov2017bimetric}] that the $\bar Q\nabla^2\dot Q$ term (which originates from the gravitational Chern-Simons term $\hat\omega d\hat\omega$ where $\hat\omega$ is the spin connection constructed from $\hat g_{ij}$) also has a coefficient completely determined by the topological properties of the quantum Hall state.  For Jain's state at large $N$, the commutator is [see also Eq.~(\ref{eq:lin}) below]
\begin{equation}\label{eq:commQQ-corr}
   [\bar Q(\x), \, Q(\x')] = \frac{64\pi\ell^2}{2N{+}1} \biggl(1+ \frac{N^2{+}N}6\ell^2\nabla^2 \biggr)\delta(\x-\x'),
\end{equation}
where we have kept only the leading and first subleading orders in the expansion over $1/N$.  At first sight, the $\nabla^2$ correction in Eq.~(\ref{eq:commQQ-corr}) does not have a corresponding counterpart in the commutators (\ref{eq:u2-comm}). However, the process of integrating out modes $u_n$ with $n\ge3$ is rather subtle:
in the Hamiltonian formalism, one cannot simply set the heavy fields $u_{\pm3}$, $u_{\pm4}$, etc., to zero in the Hamiltonian, as they do not commute with each others and with the fields $u_{\pm2}$ which are being kept.

The situation at hand is that of a Hamiltonian theory in which the phase space coordinates are divided into ``soft'' modes, denoted by $\xi_a$, and ``hard'' modes, $\xi_A$, and where there is a hierarchy between the energy scales of the soft and hard modes. At low energies, the hard modes are not excited and effectively one has the constraints $\xi_A\approx0$.  In Appendix \ref{sec:DB} we show that, if $\xi_A$ do not commute, then generally these should be considered second-class constraints, and 
the integrating out of $\xi_A$ leads to the replacement of the commutators (or Poisson brackets) $[\xi_a,\,\xi_b]$ by the Dirac brackets\cite{Dirac:1950pj,Dirac_LQM} $[\xi_a,\,\xi_b]_{\rm D}$:
\begin{equation}\label{eq:DB}
  [\xi_a,\, \xi_b]_{\rm D} = [\xi_a,\, \xi_b] - \sum_{A,B} [\xi_a,\, \xi_A][\xi,\,\xi]_{AB}^{-1}[\xi_B,\,\xi_b],
\end{equation}
where $[\xi,\,\xi]^{-1}_{AB}$ is the $AB$ component of the matrix inverse of the matrix $[\xi_A,\,\xi_B]$.  The dynamics of the soft degrees of freedom $\xi_a$ given by commuting it (using the Dirac brackets) with the effective Hamiltonian, obtained from the original Hamiltonian by setting the hard modes to zero:
 $H(\xi_a)=H(\xi_a,\xi_A)|_{\xi_A=0}$ (for more details, see Appendix \ref{sec:DB}).

In the context of our problem, the hard modes $\xi_A$ are $u_n$ with $|n|\ge3$, and the soft modes $\xi_a$ are $u_{\pm2}$. In our computation of the Dirac bracket of $u_{\pm2}$,  one needs to take into account in Eq.~(\ref{eq:DB}) only $A,B=\pm3$. Moreover, to
the order $q^2$, the inverse matrix $[u,\, u]_{AB}$ can be computed, neglecting all modes except $u_{\pm3}$.  One obtains
\begin{equation}\label{eq:DBq2}
  [u_{-2}(\x), \, u_{2}(\x')]_{\rm D} = \frac{4\pi}{2N{+}1}\biggl[ 1+ \frac{(2N{+}1)^2}{24}\ell^2\nabla^2\biggr]\delta(\x{-}\x').
\end{equation}
The coefficient in front of the $\nabla^2$ term coincides with that of Eq.~(\ref{eq:commQQ-corr}) to  subleading order in $1/N$.

\begin{figure}
  \centering
  \includegraphics[width=24em]{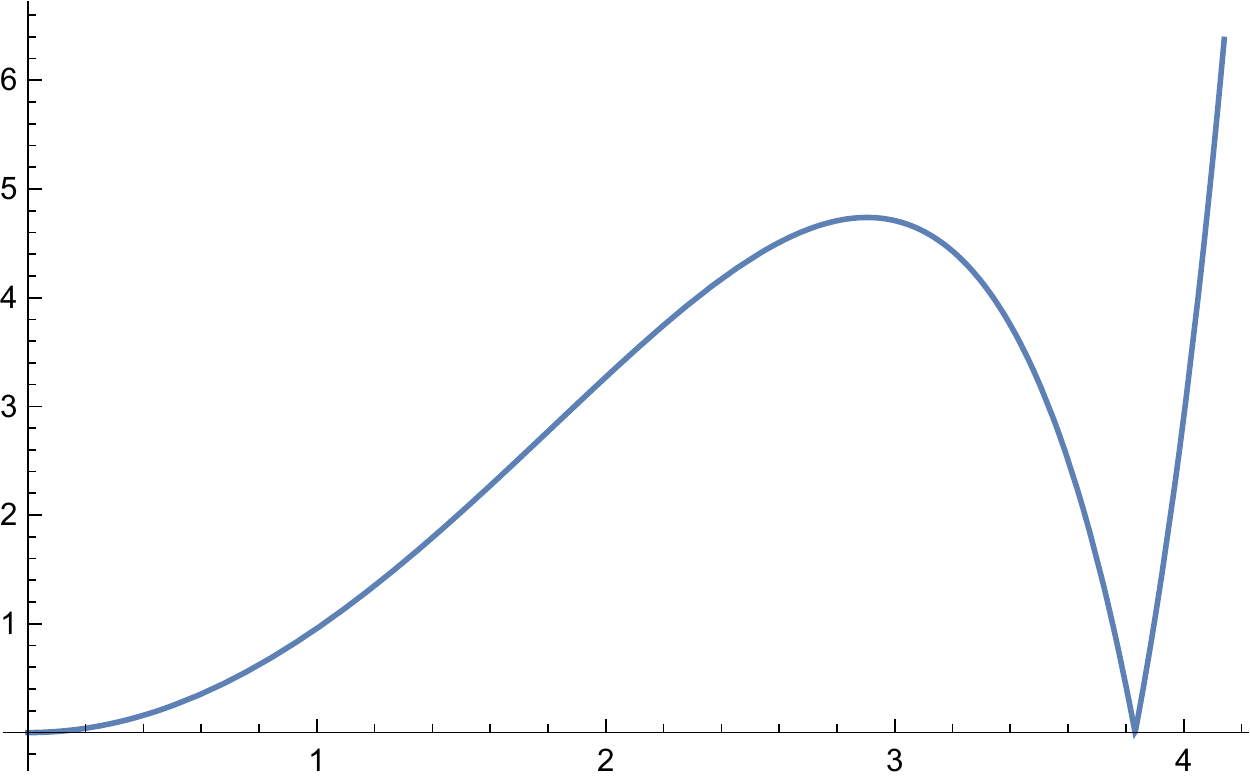}
  \caption{The shape of the dispersion curve of the neutral mode at the nematic phase transition for $\nu=N/(2N{+}1)$ in the limit of infinite $N$. The horizontal axis is $\mq=(2N{+}1)q\ell$ and the vertical axis is the energy of the excitation in arbitrary unit. The zero at $\mq\approx3.8$ should become a nonzero minimum at finite $N$.}
  \label{fig:dispersion}
\end{figure}

The $q^2$ correction in Eqs.~(\ref{eq:commQQ-corr}) and (\ref{eq:DBq2}) becomes large when $q\sim\ell/N$, i.e., when the length scale under consideration is comparable with the radius of the semiclassical orbit of a composite fermion with momentum $k_F$ moving in magnetic field $b$.
It is possible to go beyond the $q^2$ order and
evaluate the Dirac brackets between $u_{\pm2}$ from Eq.~(\ref{eq:DB}) to all orders in the momentum expansion in the regime $Nq\ell\sim1$.  The result is (see Appendix \ref{sec:evalDB} for details):
\begin{subequations}\label{eq:u2Dirac}
\begin{equation}
   [u_{-2}(\q),\, u_{2}(\q')]_{\rm D} = \frac{2\pi}{k_F^2} b F(\mq) (2\pi)^2 \delta(\q+\q').
\end{equation}
where $\mq=(2N+1)q\ell$ and $F(\mq)$ is expressed through the Bessel functions,
\begin{equation}
  F(\mq)=\frac{\mq}{2}\frac{J_1(\mq)}{J_2(\mq)}\,.
\end{equation}
\end{subequations}
The $q^0$ and $q^2$ terms in the Taylor expansion of the right-and side of Eq.~(\ref{eq:u2Dirac}) reproduces Eq.~(\ref{eq:DBq2}).
Note that $F(\mq)$ vanishes at the first zero of the Bessel function $J_1$, which has an important implication for the dispersion relation of the nematic mode. To see that, we write down the Ginzburg--Landau Hamiltonian near the nematic phase transition:
\begin{equation}\label{eq:H-nematic}
  H = \frac{k_F}{4\pi m_*}\!\int\!d^2\x\, \bigl( c_2 \ell^2 |\nabla u_2|^2 + c_0|u_2|^2 + \beta|u_2|^4) \bigr),
\end{equation}
where  $c_0$, $c_2$, and $\beta$ are phenomenological constants.  
We assume that at the nematic phase transition  $c_0$ vanishes,  but $c_2$ and $\beta$ are positive. The dispersion relation of the excitation on the symmetric side of the phase transition can be read out from Eqs.~(\ref{eq:H-nematic}) and (\ref{eq:u2Dirac}):
\begin{equation}\label{eq:disp_rel}
  \omega_2 (q) = [c_0 + c_2(q\ell)^2] \omega_c \frac{\mq J_1(\mq)}{2J_2(\mq)}\,, \quad
  \mq = (2N+1) q\ell
\end{equation}
where $\omega_c=b/m_*$.
The shape of the dispersion relation exactly at the nematic phase transition is shown in Fig.~\ref{fig:dispersion}.  At small $q$ the dispersion relation is quadratic.  Our treatment allows one to follow the dispersion curve to $q\sim N^{-1}\ell^{-1}$, where it diverges from being quadratic.  The dispersion curve reaches a maximum at $\mq\approx2.90$ and  touches zero for $\mq\approx 3.83$, the first zero of the Bessel function $J_1$.  Presumably, effects neglected in our calculation (e.g., long-ranged Coulomb interactions or terms more suppressed in powers of $N$) will transform the zero of the dispersion curve to a magnetoroton minimum.  Note that the  minimum is located at the same momentum as away from the nematic phase transition. We also find that the energy of the neutral mode rapidly increases as $\mq$ approaches the first zero of the Bessel function $J_2$, $\mq\approx5.14$. When the energy of the mode is comparable with the energy of the other excitations, the approximation breaks down and formula (\ref{eq:disp_rel}) no longer works.

Note for $q=0$ to be a local minimum of the dispersion relation, the following condition needs to be satisfied:
\begin{equation}
  c_0 < \frac{24}{(2N+1)^2}c_2 ,
\end{equation}
which, at large $N$, happens only sufficiently close to the critical point. Note also a qualitative similarity between Fig.~\ref{fig:dispersion} and the dispersion curve found in Ref.~[\onlinecite{jolicoeur2017shape}].

A general conclusion that one can draw from the above calculation is that, in the quantum Hall context, the presence of  a minimum at finite momentum in the dispersion curves of the neutral mode (a rather prevalent feature of quantum Hall state) does not necessarily signify an incipient phase transition towards an inhomogeneous phase such as a Wigner crystal.  In fact, as we have seen, as long as $c_2$ and $\beta$ are positive at the phase transition, the magnetoroton at finite $q$ never becomes unstable.  Rather, near the phase transition  another minimum develops at $q=0$, and that mode is the one that becomes unstable when the phase transition is crossed. This is consistent with Ref.~\onlinecite{regnault2016evidence}, where a nematic phase was found in a rather large part of the interaction parameter space.

\section{Bimetric theory for Jain states}
\label{sec:bimetric}

In this section, we will review the bimetric theory, developed in Ref.~[\onlinecite{gromov2017bimetric}] (see also [\onlinecite{Gromov2017Anisotropic}]) and apply it to the Jain series.

\subsection{Leading order terms}

Bimetric theory describes the gapped dynamics of a spin-$2$ degree of freedom $\hat g_{ij}$. We assume that $\hat g_{ij}$ is a symmetric, positive-definite rank-2 tensor. The inverse of $\hat g_{ij}$ will be denoted as $\hat G^{ij}$. We will refer to $\hat g_{ij}$ as intrinsic metric, meaning that it is the emergent geometric degree of freedom of the physical FQH system.

 Given this tensor we can define the vielbein according to
\be
\hat g_{ij} = \hat e^\alpha_i  \hat e^\beta_j \delta_{\alpha\beta}\,. 
\ee
Given this data we introduce the Christoffel connection $\hat \Gamma^l{}_{k,i}$ and the spin connection $\hat \omega_\mu$ through the compatibility conditions
\bea\la{dg}
\hat \nabla_k \hat g_{ij} &=& \p_k \hat g_{ij} - \hat \Gamma^l{}_{k,i}\hat g_{lj} - \hat \Gamma^l{}_{k,j}\hat g_{li}=0\,,
\\\la{de}
\hat \nabla_\mu \hat e^A_\nu &=& \p_\mu \hat e^A_\nu - \hat \Gamma^\lambda{}_{\nu,\mu} \hat e^A_\lambda + \hat \omega^A{}_{B,\mu} \hat e^B_\nu = 0\,.
\eea
Solving the compatibility conditions, Eqs.~\eqref{dg} and \eqref{de}, we find the explicit expressions for the connection
\bea
\hat \omega_0 &=& \frac{1}{2} \epsilon_\alpha{}^\beta \hat E^i_\beta \p_0 \hat e_i^\alpha\,, \la{eq:omegahat0}   \\ 
\hat \omega_j &=& \frac{1}{2} \epsilon_\alpha{}^\beta \Big( \hat E^i_\beta \p_j \hat e_i^\alpha - \hat\Gamma^k{}_{i,j} \hat e^\alpha_k \hat E_\beta^i\Big)\,, \la{eq:omegahatj}
\\
\hat\Gamma^{i}{}_{k,j}&=&\frac{1}{2}\hat G^{i\ell}\left(\partial_j \hat g_{k\ell}+\partial_k\hat g_{j\ell}-\partial_\ell \hat g_{jk}\right)\,,\label{eq:christoffelhat1}\\
\hat\Gamma^i{}_{j,0}&=&\frac{1}{2}\hat G^{ik}\partial_0 \hat g_{jk}\,.\label{eq:christoffelhat2}
\eea

In addition to the intrinsic metric $\hat g_{ij}$ we also introduce the ambient metric $g_{ij}$, and all corresponding objects without ``hats''. Given the two copies of geometric data we will allow for breaking the two corresponding copies of the diffeomorphisms down to the diagonal subgroup. This reduces the symmetry and allows for new structures, not present in a single copy of geometry. 

For the FQH application we will impose a constraint $\sqrt{g} = \sqrt{\hat g}$. This constraint leads to the removal of the ``dilaton'' degree of freedom from the dynamics. It will effectively reduce the fluctuations of $\hat g_{ij}$ down to the area-preserving distortions.

We also introduce a one-form
\be\la{eq-CDEF}
C^i{}_{j,\mu} = \Gamma^i{}_{j,\mu} - \hat \Gamma^i{}_{j,\mu}\,.
\ee
As a difference of two connections, $C^i{}_{j,\mu}$ transforms like a rank-$3$ tensor. One can construct two independent one-forms from $C^i{}_{j,k}$: the trace $C^i{}_{i,k} \sim \p_k \ln  \sqrt{\frac{\hat g}{g}} $ and the antisymmetric part $C_\mu = \epsilon_i{}^jC^i{}_{j,\mu}$.  The former vanishes due to the FQH constraint, while the latter does not.  Note, that $C_\mu$ has nice transformation properties \emph{only} when the same diffeomorphism is applied simultaneously to $\Gamma$ and $\hat \Gamma$. Thus any action that involves $C_\mu$ will break the two copies of diffeomorphisms (acting on $g$ and $\hat g$ correspondingly) down to a diagonal subgroup.

The Lagrangian for the bimetric theory, to the leading order in derivatives, is given by\cite{gromov2017bimetric}
\be\la{eq:bm}
\mathcal L_{\rm bm,1} = \frac{\nu \varsigma}{2\pi} Ad\hat \omega -\frac{\tilde m}{2}\left(\frac{1}{2} \hat g_{ij} g^{ij} - \gamma  \right)^2 - \frac{\nu \varsigma}{4\pi} (\partial_i E_i) B\,, 
\ee
where $\nu = \frac{N}{2N+1}$ is the filling fraction, $\varsigma=\frac{\mathcal{S}-1}{2}$ ($\mathcal{S}$ is the shift). The parameter $\tilde m$ sets the gap of the spin-2 mode, and $\gamma$ tunes the theory between the isotropic and nematic phases. Here we will be interested in the isotropic phase and choose $\gamma <1$. Finally, the last term appears because in the LLL the temporal component of the spin connection must enter in the combination with the electric field\cite{prabhu2017electrons}, $\hat \omega_0 + \frac{1}{2} \partial_i E_i$.

Taking a variation over $A_0$ we find that the \emph{projected} electron density is given by the Ricci curvature of $\hat g_{ij}$
\be
 \rho = \frac{\nu \varsigma}{4\pi} \hat R\,.
\ee

Finally, we note that for the Jain series
\be
\nu \varsigma = \frac{1}{8} \Bigl(2N+1 - \frac{1}{2N+1} \Bigr) \approx \frac{2N+1}{8}\,,
\ee
where in the last equation, as in most of this paper, we have kept only the leading and next-to-leading orders in $1/N$. 

\subsection{Higher gradient corrections}

In what follows we will need to consider terms that are higher order in the gradient expansion than the terms we described so far. The two sub-leading terms of interest\cite{gromov2017bimetric} are the (non-universal) kinetic energy term and the (universal) gravitational Chern-Simons term. These terms must emerge on the general grounds as was discussed in detail in Ref.~[\onlinecite{gromov2017bimetric}]. 

We start with the non-universal term
\begin{multline}\la{eq:Skin}
S_{\rm kin}[\hat g;g] = - \frac{\alpha}{4}\! \int\! d^3x\, \sqrt{g}~  g^{kl} C_{k} C_{l} \\ \sim -\frac{\alpha}{4} \!\int\! d^3 x\, \sqrt{g} \left| \Gamma - \hat \Gamma \right|^2\,.
\end{multline}
The geometric meaning of \eqref{eq:Skin} was discussed in great detail in Ref. [\onlinecite{gromov2017bimetric}]. Close to the nematic phase transition point this term must be added to the action. It will contribute to the $q^2$ behavior of the GMP mode dispersion relation. We will discuss the effects of this term quantitatively in the following sections.

The subleading universal term was also discussed in great detail in Ref.~[\onlinecite{gromov2017bimetric}]. It is a purely gravitational Chern-Simons term
\be\la{eq:gcs}
 S_{\rm gCS}[\hat g] = -\frac{\hat c}{4\pi} \!\int \! \hat \omega d \hat \omega - \frac{\hat c}{8\pi} \ell^2 \!\int\!  \p_i E^i \hat R \,,
\ee
The coefficient of the $\p_i E^i \hat R$ term is discussed in Ref.\,[\onlinecite{gromov2017bimetric}].

The effects of the term are subtle, but curious. We refer the interested reader to the Ref.\,[\onlinecite{gromov2017bimetric}]. Here we will only be interested in the linearization. The gravitational Chern-Simons term can be expanded as
\be
\int\! \hat \omega d \hat \omega = \int\! \hat \omega_0\hat R - \epsilon_{ij} \hat \omega_i \dot{\hat \omega}_j\,.
\ee
Note that the first term is inherently non-linear and its effects are presently not understood. 
The coefficient $\hat c$ has been computed for the Jain states previously \cite{gromov2017bimetric} and is given by
\begin{equation}\label{eq:hatc}
\hat c \approx \frac{N^2(2N+3)}{24}\,.
\end{equation}

The full bimetric Lagrangian is given by
\be\la{bmfull}
\cL_{\rm bm}[\hat g] = \cL_{\rm bm,1}[\hat g] + \cL_{\rm gCS}[\hat g] + \cL_{\rm kin}[\hat g]\,.
\ee

\subsection{Linearized effective Lagrangian}

To analyze the theory (and to compare it to the DCF) it is convenient to linearize it in flat space -- setting $g_{ij} = \delta_{ij}$ -- close to the minimum of the potential (second term in Eq. \eqref{eq:bm}). Then the intrinsic metric can be parametrized as in Eq.~(\ref{hatgQ}). We linearize Eqs.\,\eqref{hatgQ},\eqref{bmfull} close to $Q_i=0$. Putting everything together we find the linearized theory
\begin{multline}\label{eq:lin}
\cL_{\rm bm} =  - \frac{i}{2} \frac{2N{+}1}{32\pi } \biggl[ \ell^{-2} \bar Q \dot { Q} -  \frac{1}{2}\Big( E \p \bar Q - \bar E \bar \p  Q\Big) \biggr] 
 \\ 
 - \frac{i}{2} \frac{N^2(2N{+}3)}{192 \pi} \biggl[ \bar Q \Delta \dot Q + \frac{1}{4}\Big(\bar E \bar \p \Delta Q + E \p \Delta \bar Q 
 \\
 + \bar E \bar \p^3 Q + E \p^3 \bar Q \Big)\biggr] -  \frac{m}{2} Q \bar Q -\alpha  |\partial Q|^2 ,
\end{multline}
where  $ m = (1-\gamma) \tilde m$ and we have used the linearized Ricci curvature $\hat R \approx 2i (\p^2 \bar Q - \bar \p^2 Q) $.

The commutator between $Q$ and $\bar Q$ can be read from the $\bar Q\dot Q$ and  $\bar Q\Delta\dot Q$ terms in Eq.~(\ref{eq:lin}); we find Eq.~(\ref{eq:commQQ-corr})

\subsection{Poisson structure}
Upon canonical quantization of \eqref{eq:bm} we find the following canonical commutation relations (CCRs)
\be
[\hat E^i_\alpha(\bx), \hat e^\beta_j(\bx^\prime)] = -\frac{2i}{ \rho_0 \varsigma}\delta^i_{j}\epsilon_{\alpha}{}^{\beta}\delta(\bx - \bx^\prime),
\ee
where $\rho_0 = \frac{\nu}{2\pi \ell^2}$.
These CCRs lead to the following commutation relations for the metric
\begin{multline}
[ \hat g_{ij}(\bx), \hat g_{kl}(\bx^\prime)] = -\frac{2i}{ \rho_0 \varsigma}\big( \epsilon_{il} \hat g_{jk} + \epsilon_{jk}\hat g_{il} \\
 + \epsilon_{jl}\hat g_{ik} + \epsilon_{ik}\hat g_{jl}\big)  \delta(\bx - \bx^\prime),
\end{multline}
which coincides with the $\mathfrak{sl}(2,\mathbb R)$ Lie algebra (up to rescaling).
The mapping onto the $\mathfrak{sl}(2,\mathbb R)$ algebra becomes evident in complex coordinates:
\begin{subequations}
\begin{align}\la{eq:CCR1}
&[\hat g_{zz}(\bx),\, \hat g_{\bar z \bar z}(\bx^\prime)] = \frac{16}{\rho_0 \varsigma} \hat g_{z\bar z}(\bx)\, \delta(\bx - \bx^\prime)\,,\\
&[\hat g_{\bar z\bar z}(\bx),\, \hat g_{ z \bar z}(\bx^\prime)] = \frac{8}{\rho_0 \varsigma} \hat g_{\bar z\bar z}(\bx) \delta(\bx - \bx^\prime)\,,
\\ \la{eq:CCR3}
&[\hat g_{zz}(\bx),\, \hat g_{ z  z}(\bx^\prime)]= [\hat g_{ z\bar z}(\bx),\, \hat g_{ z \bar z}(\bx^\prime] =0\,.
\end{align}
\end{subequations}
In terms of the complex variables $Q$ and  $\bar Q$, the components of the metric tensor are found to be
\bea\la{eq:gzzLIN}
&&g_{zz} =- 2i Q \frac{\sinh\sqrt{Q\bar Q}}{\sqrt{Q\bar Q}}\approx -2 i  Q, \\ \la{eq:gzbLIN}
&& g_{ z \bar z} = 2 \sqrt{Q \bar Q} \approx 2 - Q\bar Q .
\eea
Eqs. \eqref{eq:CCR1}--\eqref{eq:gzbLIN} imply that, at the linearized level 
\be
[Q(\bx),\bar Q(\bx^\prime)] = - \frac{4i}{\rho_0 \varsigma} \delta(\bx - \bx^\prime),
\ee
At nonlinear level, the fields $Q$ and $\bar Q$ do not have simple commutation relations, however, the components of the intrinsic metric tensor, which are nonlinear functions of $Q$ and $\bar Q$, do have simple CCRs. We will only require the linearized CCRs for the composite Fermi liquid theory discussed in the next section.

\section{Dirac Composite Fermi Liquid theory}
\label{sec:DCFL}

In this section, we briefly review the Dirac composite Fermi liquid theory; in particular, we introduce the action and fix the notations. The full derivation of the model can be found in Refs.~[\onlinecite{NS}],[\onlinecite{NGRS}]. 
 By extending the Dirac composite Fermi liquid theory to a finite background magnetic field, one can describe the Jain states $\nu_\pm=\frac{1}{2}\pm \frac{1}{2(2N+1)}$ near the half filling, where $N$ is large.

 To simplify the presentation, in this section, we will use an alternative notation for the complex components of a vector
\be
V_\pm = \frac{V_1 \mp i V_2}{2}\,.
\ee

\subsection{Dirac composite fermion}

Here we summarize the result of Ref.~[\onlinecite{NGRS}].  For simplicity, we will concentrate on the case of short-ranged interaction, for which the results can be written in more compact form.

The action of the Dirac composite fermion theory, which can be used to obtain results to leading and next-to-leading order in $1/N$, is
\begin{multline} \label{eq:Dirac2}
S[\psi,  a;\mathcal A]= i\! \int\!d^3x \,  \bigl(\psi^\dagger D_t \psi+    v_F \psi^\dagger\sigma^i D_i \psi \bigr)\\
   + \frac{1}{4\pi}\! \int\!d^3x\, \Bigl( -\tilde A da + \frac12\cA d\cA - \frac{m_*}{2B} 
   \tilde E_i \tilde E_i\Bigr) ,
\end{multline}
where $\mathcal A_i = \tilde A_i=A_i$, 
\begin{align}
  \cA_0 & = A_0 + \frac{\nabla \cdot E}{4B}\,,\\
  \tilde A_0 &= A_0 + C_0\delta B,
\end{align}
and $C_0$ is a constant determined by the potential $V(r)$ of the two-body interaction between the electrons,
\begin{equation}
 C_0 = \frac12\!\int\limits_0^\infty\!dr\, r\biggl[ 1-\biggl( 1-\frac{r^2}{2\ell^2} \biggr) e^{-r^2/2\ell^2} \biggl] V(r).
\end{equation}

As explained in Ref.~[\onlinecite{NGRS}], the kinetic terms for $a_\mu$, generally allowed by symmetries, can be neglected if one works to leading and next-to-leading orders in $1/N$.  The absence of the kinetic term for $a_\mu$ leads to the following constraints 
\begin{subequations}\label{eq:constraints_Dirac}
\begin{align}
  \rho_{\rm CF} &= \frac B{4\pi} \,,\\
  j_{\rm CF}^i  &= \frac1{4\pi}\epsilon^{ij} \tilde E_j,
\end{align}
\end{subequations}
where $E_j=\d_j \tilde A_0 - \d_0 A_j$.

\subsection{Bosonized approach}

In this section, we employ the bosonized approach \cite{golkar2016higher,NS,NGRS} to study the theory of Dirac CF in the long wavelength limit where the RPA and the Fermi liquid are equivalent \cite{nguyen2016exact} up to next-to-leading order in $1/N$.  

\subsubsection{Shape quantization of Fermi surface}

When the emergent magnetic field $b$ is weak, one can start from a composite Fermi liquid with Fermi momentum $k_F$. The dynamics of the composite fermions at low energy regime can be parametrized by bosonic perturbations of the shape of the Fermi surface, Eq.~(\ref{eq:FS-shape}).
The composite fermion density can be written in terms of the dilation of the FS, $u_0$, as \cite{nguyen2016exact}
\begin{equation}
\label{eq:unrho}
\rho_{CF}=\bar{\rho}_{CF}+\frac{k_F}{2\pi}u_0\,,
\end{equation}
while the CF current, $J_{CF}$, can be written in terms of $u_{\pm 1}$ \cite{nguyen2016exact} 
\be
\label{eq:unP}
\!\!\!\!J_{CF}^{\pm}=\frac{k_F v_F}{4\pi}u_{\pm 1}\,.
\ee 
In the Fermi-liquid approximation the momentum density of the CFs is  $P^{\pm}_{CF}=\frac{k_F}{v_F}J^{\pm}_{CF}$.

Note that $u_0$ is proportional to the fluctuation of the density of the CFs, and $u_{\pm1}$ to the current.  We can identify $u_1=u_{\bar z}$ and $u_{-1}=u_z$, then $j_i\sim u_i$, with $i$ being the spatial index. The composite fermion theory is now formulated in terms of a Hamiltonian
$H = H(u_n)$
with the  canonical commutation relations (CCR) (\ref{eq:commu})\cite{golkar2016higher,NGRS}

For the Hamiltonian, one can take the following quadratic functional \cite{golkar2016higher,NGRS} 
\begin{equation}
\label{eq:HamCF}
H_{CF}=\frac{v_Fk_F}{4\pi}\!\int\! d^2 \mathbf{x}\!\!
   \sum_{n=-\infty}^{\infty}(1+F_n)u_n(\mathbf{x})u_{-n}(\mathbf{x}),
\end{equation}
where $F_n$ are the Landau parameters, and $g$ a gauge coupling constant. We take $F_1=0$, since $1+F_1$ can be absorbed in the normalization of the Fermi-velocity $v_F$.  Note that $F_{n}=F_{-n}$, by the invariance with respect to the product of time reversal and spatial reflection.  For convenience, we will use the notation $f_n=1+F_n$.

Near the nematic phase transition, $F_2 \to -1$ and the $u_{\pm2}$ modes become light.  In this case a gradient term $\nabla u_2\nabla u_{-2}$ needs to be added to the Hamiltonian.  The modification is equivalent to replacing the Landau parameter $F_2$ by a momentum-dependent Landau parameter, i.e.,
\begin{equation}\label{eq:f2}
  f_2 = c_0+c_2 (q\ell)^2,
\end{equation}
where we assume 
\begin{equation}
c_0 \ll 1, \qquad c_2 (q\ell)^2 \ll 1.
\end{equation} 
We also assume that $f_n= O(1)$ for $n>2$. In the low energy limit $\omega/\omega_c \ll 1$, $u_n$ ($|n|>2$) modes can be integrated out leading the single-mode effective field theory described below.

\subsubsection{Equations of motion and constraints}


The equations of motion for $u_n$, coming from the Eqs.\eqref{eq:commu},\eqref{eq:HamCF} are
\begin{equation}
\label{eq:Heisenberg}
\frac{d u_n}{ dt}=-i [u_n,H]+\frac{\partial u_n}{\partial t}.
\end{equation}
The last term in Heisenberg equation \eqref{eq:Heisenberg} is nonzero whenever $u_n$ explicitly depends on time. The explicit form of $u_{\pm 1}$ is given by
\be\la{eq:upmexpl}
u_{\pm 1}=-i\frac{4\pi }{k_F^2}\psi^\dagger \tilde{\cD}_{\pm}\psi\,.
\ee
Applying $\p_t$ to both sides we find, at  linear order in perturbations,
\be
\partial_t u_{\pm1}=-\frac{4\pi }{k_F^2} \partial_t (\delta a_\pm) \psi^\dagger \psi
=-\partial_t \delta a_{\pm}\,.
\ee
A similar argument \footnote{We use the explicit form of $u_{\pm n}$, for $n>0$, as $u_n\sim \psi^\dagger (\tilde{D}_+)^n\psi$, and $u_{-n}\sim \psi^\dagger (\tilde{D}_-)^n\psi$} leads to
\begin{equation}
\label{eq:unexpl}
\partial_t u_{n}=0, \qquad (n\neq \pm 1).
\end{equation}
to linear order in perturbations.
Using Hamiltonian \eqref{eq:HamCF} together with CCR \eqref{eq:commu} and Eqs. \eqref{eq:upmexpl} and \eqref{eq:unexpl}, we arrive at
\bea
\nonumber
	0&=&\Big[\omega+\sign(\bar{b})nf_n\omega_c\Big]u_n(\mq,\omega)
	\\
	&+&\omega\Big(\delta_{n,1}\delta \tilde{a}_++\delta_{n,-1}\delta \tilde{a}_{-}\Big)\nonumber
	\\ \label{eq:recur1}
	&-&\omega_c\Big[\mq_{\bar{z}}f_{n+1}u_{n+1}(\mq,\omega)+\mq_z f_{n-1}u_{n-1}(\mq,\omega)\Big],
\eea
where 
\be
\omega_c=\frac{|\bar b|v_F}{k_F}\,, \qquad \mq_i=q_i \ell\frac{\bar{B}}{|\bar b|} = (2N+1)q_i \ell \,,
\ee
and $f_n = 1+F_n$, and $\mq_z$ and $\mq_{\bar z}$ are defined according to the conventions in Appendix (\ref{sec:conventions}). 
In the derivation of \eqref{eq:recur1}, we have ignored the terms $\delta \tilde{a}\cdot u_n$, since they are higher order in perturbation.

The constraints~(\ref{eq:constraints_Dirac}), in terms of the bosonic variables, read
\be \la{eq:constr2}
u_0=\frac{\delta B}{2k_F}\,, \quad u_{\pm 1} = \pm i\frac{\tilde E_\pm}{(2N+1)\omega_c}\,.
\ee
These constraints will prove crucial in what follows.
\subsection{DCF liquid near a nematic phase transition}
\label{sec:SMA}

In this section, we derive the effective action for the DCF liquid close to a nematic phase transition, where $f_2=c_0 \ll 1$. Near the phase transition, we should add an extra contribution $\sim u_2 \nabla^2 u_{-2}$ to the Hamiltonian \eqref{eq:HamCF}, since the term $f_2u_2 u_{-2}$ is no longer dominant. 

\subsubsection{Effective Lagrangian}

The recursion relation \eqref{eq:recur1} for $|n|>2$ for the Jain series at filling $\nu_-$ (when $\sign(\bar{b})=1$), at small frequency takes form
\be \la{eq:unsol}
u_{\pm n} = \left( \pm \frac{2\mq_{ \pm}}{\mq}\right)^{\pm n-2}\frac{J_n(\mq)}{J_2(\mq)}\frac{f_2}{f_n}u_{\pm 2}, \qquad |n|>2
\ee

Using Eq.\,\eqref{eq:unsol} the remaining Eqs.\,\eqref{eq:recur1} are brought to simple form \footnote{We have also assumed $\delta B =0$, setting $u_0 = 0$} 
\bea
\!\!\!\!\!\!\!\!\!\!\!\! (\omega \pm \omega_c)u_{\pm 1}+ \omega \delta a_{\pm} - f_2\omega_c \mq_{\mp}u_{\pm 2} = 0\,, \,\,\,\,\, n=\pm1, 
\\
\!\!\!\!\!\!\!\!\!\!\!\!\Big(\omega \pm f_2 \omega_c F(\mq) \Big) u_{\pm 2} - \omega_c\mq_\pm u_{\pm 1} = 0\,, \,\,\,\,\, n = \pm 2, \label{eq:u2F}
\eea
and  
\be
F(\mq)=2-\frac{\mq J_3(\mq)}{2 J_2(\mq)}=\frac{\mq}{2}\frac{J_1(\mq)}{J_2(\mq)}\,.
\ee


Notice that Eq.~(\ref{eq:u2F}) can be obtained by using the Hamiltonian (\ref{eq:HamCF}), where all terms involving $u_n$ with $|n|\ge3$ have been dropped, with the Dirac bracket Eq.~(\ref{eq:u2Dirac})
Expanding Eq.~(\ref{eq:u2Dirac}) over $q$, we find, to order $q^2$, Eq.~(\ref{eq:DBq2}).  Since the computation of the Dirac brackets do not involve the Hamiltonian, the dynamics at low energy is independent of the higher Landau parameters $F_n$ with $|n|\ge3$.

Returning to the equations of motion, we now use the constraints \eqref{eq:constr2} to exclude $u_{\pm 1}$ and determine the gauge field $a_\pm$ in terms of $u_{\pm 2}$. We find
\bea
\label{eq:rec1}
&&\omega \delta a_{\pm} = \mp i \frac{\omega \pm \omega_c}{(2N+1)\omega_c}\tilde E_{\pm} -\omega_c q_{\mp} f_2 u_{\pm 2}\,, 
\\ \la{eq:u2fin}
&&\Big[\omega \pm \omega_2(\mq) \Big] u_{\pm 2} = \pm  i \ell q_\pm  \tilde E_{\pm}\,,
\eea
where
\begin{equation}
  \omega_2(\mq) = f_2 \omega_c F(\mq).
\end{equation}  

According to Eqs.~(\ref{eq:u2fin}), $\omega_2(\mq)$ is the frequency of oscillations of $u_{\pm2}$ in the absence of external sources, and hence is the gap of the spin-2 mode. 
 
Thus, we have shown that every degree of freedom is completely determined by the quadrupolar deformation of the Fermi surface $u_{\pm 2}$, which, in turn, obeys Eq.~\eqref{eq:u2fin}. We wish to find a variational principle for Eq.~\eqref{eq:u2fin}, which is easy to do since the equation is linear in $u_{\pm 2}$.  Fixing the coefficient of the $adA$ term to be the same as in Eq.~(\ref{eq:Dirac2}), we find the Lagrangian to be
\begin{multline}
\label{eq:SMA}
\cL^{\rm SMA}_{\nu_-} = \frac{2N+1}{2\pi F(\mq)}\big[ \omega - \omega_2(\mq) \big]u_2(-\omega,-\mathbf{q})u_{-2}(\omega,\mathbf{q})
\\
-\frac{(2N+1)\ell}{2\pi F(\mq)}\Big[iq_- \tilde E_{-}(-\omega,-\mathbf{q})u_2(\omega,\mathbf{q}) 
\\
+  iq_{+} \tilde E_{+}(-\omega,-\mathbf{q}) u_{-2}(\omega,\mathbf{q})\Big] , 
\end{multline}
The full effective Lagrangian is obtained after integrating out $\delta a_\mu$, which generates a number of contact terms that we denote collectively as $\cL^{\rm ct}_{\nu_-}[A_\mu]$. The full effective Lagrangian is thus given by
\be\la{eq:LagSMA1}
\cL_{\nu_-} = \cL^{\rm SMA}_{\nu_-}[u_{\pm 2}; A_\mu] + \cL^{\rm ct}_{\nu_-}[A_\mu]\,,
\ee
where 
\bea\nonumber
\label{eq:cont}
\cL^{\rm ct}_{\nu_-}[A_\mu] &=& i\frac{\nu_-}{4\pi} \epsilon^{\mu\nu\rho}A_\mu(-\omega,-\mathbf{q}) q_\nu A_\rho(\omega,\mathbf{q})\\
&-&\frac{\ell^2}{16\pi}\left[1-\frac{2N+1}{F(\mq)}\right]iq_kE^k(-\omega,-\mathbf{q})B(\omega,\mathbf{q})\nonumber.\\
\eea
Detailed derivation of \eqref{eq:LagSMA1} is left for the Appendix \ref{sec:deriSMA}.

The positive Jain series can be obtained following the same steps, but setting $\sign(\bar b) = -1$ in \eqref{eq:recur1}. 

\subsubsection{Mapping between DCF and Bimetric theory}

The goal of this section is to establish precise dictionary between the effective Lagrangians \eqref{eq:lin} and \eqref{eq:LagSMA1} in their common regime of validity, i.e., large $N$ and $q\ell N\ll1$.
This mapping provides a non-linear completion of \eqref{eq:LagSMA1} and a procedure of coupling the quadrupolar fluctuations of the Fermi surface to the ambient geometry in this regime.
  
 To leading and next-to-leading orders in the gradient expansion, we can replace $F(\mq)=2 - \frac{\mq^2}{12}$ in Eq.~\eqref{eq:LagSMA1} to find 
\begin{multline}\label{eq:SMAdyn}
  \cL^{\rm SMA}_{\nu_\pm} = \mp \frac{i}{2}\frac{2N+1}{2\pi}\left[ u_2 \dot u_{-2}+i\ell(\tilde E_- \bar \partial u_2 + \tilde E_+\partial u_{-2} )\right] 
\\
\pm \frac{i}{2}\frac{N^2(2N{+}3)\ell^2}{12\pi}\Big[ u_2 \Delta \dot u_{-2} 
+i\ell(\tilde E_- \bar \partial\Delta u_2 + \tilde E_+\partial\Delta u_{-2} )\Big]
\\  
-\frac{(2N+1)\omega_c}{2\pi}u_2 (c_0 - c_2 \ell^2 \Delta) u_{-2} \,.
\end{multline}
We have written this action for both Jain sequences.
 
 Comparing Eq.~\eqref{eq:lin} with Eq.~\eqref{eq:SMAdyn}, we find that the DCF and bimetric actions are equivalent upon the following identifications of parameters:
\bea
  \label{eq:map1}
  \bar Q &=& -4i\ell u_{-2}\,, \quad Q = 4i\ell u_{2},
\\
    \label{eq:map2}
m&=&\frac{c_0(2N+1)\omega_c}{16\pi \ell^2}\,, \quad \alpha=\frac{c_2(2N+1)\omega_c }{32\pi}.
\eea
We have matched three parameters: $m$, $\alpha$ and the relative coefficient between $u_{\pm 2}$ and $Q$. As a consequence of the matching the coefficients of the gravitational Chern-Simons term (in its linearized version) was matched automatically. This matching correspondence can be traced to the fact that both theories produce the same projected static structure factor in leading and sub-leading orders in momentum expansion.

The map between DCF and bimetric theory can be considered as an alternative derivation of the bimetric theory.  In the next section we will compute the  linear response functions. These computations, at the tree level, can be done in either theory with identical results.

\section{Linear response in DCF theory}
\label{sec:linres_DCF}

In this section we will compute the linear response from the effective Lagrangian \eqref{eq:LagSMA1}. 

\subsection{Jain state $\nu_-$}
Integrating out $u_{\pm 2}$ reduces to the substitution of
\be
u_{\pm 2}=\pm\frac{i \ell q_\pm \tilde E_\pm}{\omega\pm f_2 \omega_c F(\mq)},
\ee
back into the Lagrangian \eqref{eq:LagSMA1}. This leads to the effective action for the external electromagnetic field $S[A_\mu]$. Below we present the linear response functions that follow from that action.

The charge density is given by
\begin{equation}
\rho_{\nu_-}=\frac{N}{2N+1}\frac{\bar B}{2\pi},
\end{equation}
and corresponds to the filling factor $\nu=N/(2N+1)$ in a Jain's sequence. 

The susceptibility is given by
\begin{equation}
\chi_{\nu_-}(\omega, q)=\frac{\mq^4}{16\pi (2N+1)^3\ell^2}\frac{f_2\omega_c}{\omega^2-\omega_2^2(q)}.
\end{equation}

The projected static structure factor, defined as,
\begin{equation}\la{eq:SSF}
\bar{s}_\nu=-\frac{i}{\rho_\nu}\!\int\! \frac{d\omega}{2\pi}\,\chi_\nu(\omega,q),
\end{equation}
takes the simple form independent of $f_2$,
\begin{equation}
\nu_{\pm}\bar{s}_{\nu_{\pm}}=\frac{q^3\ell^3}{8}\frac{J_2(\mq)}{J_1(\mq)}
\end{equation}
To compare the result with the bimetric theory, one expands the result in a double series over  $q$ and $1/N$.  One can see that, to leading and next-to-leading orders in $1/N$, the static structure factor, up to $q^6$ order, can be written as
\begin{equation}
\label{eq:SSFp}
\nu_-\bar{s}_{\nu_-}(q)=\frac{\nu_-(\cS_{\nu_-}{-}1)}{8}(q\ell)^4+\frac{\hat{c}}{8}(q\ell)^6+\ldots,
\end{equation}
where $\cS_{\nu_-}=N+2$ is the Wen-Zee shift of the Jain state at filling factor $\nu_-$ and $\hat c$ was defined in Eq.~(\ref{eq:hatc}). 

For the Hall conductivity, it is more convenient to write formulas for
\begin{equation}
  \tilde\sigma^H(\omega, q) = \sigma^H(\omega,q) - C_0 \chi(\omega,q),
\end{equation}
which is given by, after a simple calculation,
\begin{multline}\label{eq:sigma-closed}
       \tilde \sigma^H_{\nu_-}(\omega, q)=\frac{\nu_-}{2\pi}-\frac{1}{8\pi}\left[-(q\ell)^2 +q\ell\frac{J_2(\mq)}{J_1(\mq)}\right]\\-\frac{1}{8\pi (2N+1)^2q\ell}\frac{J_2(\mq)}{J_1(\mq)}\frac{\omega^2}{\omega^2-\omega_2^2(q)} \,.
\end{multline}
To order $q^2$, $\sigma^H$ and $\tilde\sigma^H$ coincide and is given by
\begin{multline}\label{eq:Hallp}
  \sigma^H_{\nu_-}(\omega,q)=\frac{\nu_-}{2\pi}\left[1+\frac{ \cS_{\nu_-}-2}{4}(q\ell)^2\right]\\
  +\frac{(2N+1)(q\ell)^2}{32\pi}\frac{\omega^2}{\omega^2-\omega_2^2(q)} +\cO(q^4)\,.
\end{multline}
 We define the ``projected Hall conductivity'' by 
\begin{equation}
\bar{\sigma}^H_\nu=\sigma^H_\nu-\nu \sigma^H_1,
\end{equation}
where
\begin{equation}
\sigma_1^H=\frac{1}{2\pi}\left[1-\frac{1}{4}(q\ell)^2\right]
\end{equation}
 is the Hall conductivity of a completely filled Landau level.  The projected DC Hall conductivity is given by 
\begin{equation}
\bar{\sigma}^H_{\nu_-}(0,q)=\frac{\nu_-}{2\pi}\frac{\cS_{\nu_-}{-}1}{4}(q\ell)^2+\cO(q^4).
\end{equation}
We find that the bimetric theory~\eqref{eq:lin}, where the contact terms have been discarded, evaluates the projected Hall conductivity directly.

\subsection{Jain state $\nu_+$}
Performing a similar analysis for Jain state with filling factor $\nu_+=(N+1)/(2N+1)$, one obtains the susceptibility
\begin{equation}
\chi_{\nu_+}(\omega, q)=\frac{\mq^4}{16\pi (2N+1)^3\ell^2}\frac{f_2\omega_c}{\omega^2-\omega_2^2(q)} \,.
\end{equation}
From now on we will work to order $q^6$ in the structure factor and $q^2$ in the Hall conductivity. The static structure factor is
\begin{equation}
\label{eq:SSFh}
\nu_+\bar{s}_{\nu_+}(q) = -\frac{\nu_+(\cS_{\nu_+}{-}1)}{8}(q\ell)^4+\frac{\hat{c}}{8}(q\ell)^6 + \cO(q^8),
\end{equation}
where the shift for $\nu_+$ is $\cS_{\nu_+}=-N+1$. 

 The relation $\nu_+ \bar s_{\nu_+} = \nu_- \bar s_{\nu_-}$ holds \cite{nguyen2016particle} as can be clearly seen from Eq.~\eqref{eq:SSFp} and Eq.~\eqref{eq:SSFh}. 
 The ac Hall conductivity is given by
 \begin{multline}
 \label{eq:sigma+closed}
 \tilde{\sigma}^H_{\nu_+}(\omega,q)=\frac{\nu_+}{2\pi}-\frac{1}{8\pi}\left(-(q\ell)^2 -q\ell\frac{J_2(\mq)}{J_1(\mq)}\right)\\+\frac{1}{8\pi (2N+1)^2q\ell}\frac{J_2(\mq)}{J_1(\mq)}\frac{\omega^2}{\omega^2-\omega_2^2(q)},
 \end{multline}
 and to order $q^2$
\begin{multline}\label{eq:Hallh}
  \sigma^H_{\nu_+}(\omega,q)=\frac{\nu_+}{2\pi}\left[1+\frac{ \cS_{\nu_+}-2}{4}(q\ell)^2\right]\\ 
  -\frac{(2N+1)(q\ell)^2}{32\pi}\frac{\omega^2}{\omega^2-\omega_2^2(q)}+\cO(q^4).
\end{multline} 
 The projected dc Hall conductivity is given by 
 \begin{equation}
 \bar{\sigma}^H_{\nu_+}(0,q)=\frac{\nu_+}{2\pi}\frac{\cS_{\nu_+}{-}1}{4}(q\ell)^2+\cO(q^4).
 \end{equation}
Combining \eqref{eq:Hallp} and \eqref{eq:Hallh}, we arrive at the following relationship for the projected AC Hall conductivity
\begin{equation}
\bar{\sigma}^H_{\nu_-}(\omega,q)+\bar{\sigma}^H_{\nu_+}(\omega,q)=0\,,
\end{equation}
which is in agreement with identity derived in Ref.~[\onlinecite{SL}].


\section{Higher spin fields and the SMA}
\label{sec:spin3}

In this section, we address the question of the validity of the SMA for the static structure factor. For this we do not assume that the system is near the nematic phase transition and compute the static structure factor up to the $q^6$ order. To this order, only the spin-2 and spin-3 modes contribute, and the result can be written as \cite{NGRS} 
\begin{multline}\label{eq:susceptu3}
\chi(\omega,q)= - \frac{2N+1}{32\pi \ell^2}  \frac{\tilde{\omega}_2(q)}{\omega^2-\tilde{\omega}^2_2(q)} (q\ell)^4
\\ +  \frac{\hat{c}}{8\pi \ell^2 }\left[ (1-x)\frac{\Delta_2}{\omega^2-\Delta_2^2}+x\frac{\Delta_3}{\omega^2-\Delta_3^2}\right] (q\ell)^6\,,
\end{multline}
where $\tilde{\omega}_2(q)$ is the dispersion relation for the spin-2 mode,
\begin{equation}
\frac{\tilde{\omega}_2(q)}{\Delta_2} = 1-\left[-\frac{c_2}{c_0}+\frac{(2N+1)^2}{24\left(1-\frac{\Delta_2}{\Delta_3}\right)}\right](q\ell)^2+\cdots,
\end{equation}
while $\Delta_2=2 c_0\omega_c$ and $\Delta_3=3f_3\omega_c$ are the energies of the spin-2 and spin-3 modes, respectively, and
\begin{equation}
  x = \frac{\Delta_2^2}{(\Delta_3-\Delta_2)^2} \,.
\end{equation}

Next we evaluate the projected static structure factor defined in Eq.\,\eqref{eq:SSF}. Note that the only contribution to $\bar s_4$ comes from the spin-$2$ mode, while $\bar s_6$ receives the contributions from both spin-$2$ and spin-$3$ modes. More concretely, we find
\begin{equation}
\nu_{\pm}\bar{s}_{\pm}(q)=\bar{s}^{(2)}_4 (q\ell)^4+\Big(\bar{s}^{(2)}_6 +\bar{s}^{(3)}_6\Big) (q\ell)^6+O(q^8),
\end{equation}
where $\bar{s}_m^{(n)}$ is the contribution of the spin-$n$ mode to the coefficient in front of $q^m$ in the long-wave expansion of the static structure factor. Using Eq.~\eqref{eq:susceptu3} we find these coefficients 
\bea
\!\!\!\!\!	\bar{s}^{(2)}_4 = \frac{2N+1}{32}\,, \quad \bar{s}^{(2)}_6=\frac{\hat{c}}{8}\left(1-x\right)\,, \quad \bar{s}^{(3)}_6=\frac{\hat{c}}{8}x\,.
\eea
The two contributions to $\bar s_6$ are not quantized separately, but the sum is quantized.

Near the nematic phase transition, $\Delta_2/\Delta_3\rightarrow 0$ the entire contribution to the $\bar s_6$ comes from the second mode since $x\rightarrow 0$. In this case (and only in this case) the contribution of the second mode is quantized
\begin{equation}
\lim_{\Delta_2/\Delta_3\rightarrow 0}\bar{s}^{(2)}_6=\frac{N^2(2N+3)}{192}=\frac{\hat{c}}{8}\,.
\end{equation}
Thus, we confirm that the SMA becomes more and more reliable for the $q^6$ part of the SSF when one approaches the nematic phase transition.

\section{Discussions and Conclusions}
\label{sec:concl}

We have performed a detailed comparison between the bimetric and DCF theories of Jain states, close to half filling. We find that at the linear level the two theories agree in the common domain of validity: close to the nematic phase transition and at sufficiently small momentum. It would be interesting to extend the analysis of the DCF theory to nonlinear level 
and compare it to bimetric theory.  This is likely to be important if one is interested in the symmetry broken phase, which is not considered in this paper.

Bimetric theory has been aplied to the general Jain series at filling $\nu = \frac{N}{2pN+1}$. The composite fermion description of these states with $p>1$ is less certain than for those with $p=1$, where particle-hole symmetry provides a guiding principle. It was conjectured that these states are described by a FS with $\pi/p$ Berry phase \cite{Wang2016,You:2017tvv}. Perhaps bimetric theory can shed light on the composite fermion theory for $p>1$.

The DCF theory naturally comes equipped with an infinite number of higher spin fields. Our computations (in particular, those of the Dirac brackets) explicitly show that the rigid structure of the bimetric theory is only reached upon integrating out these fields.
In particular, the $q^2$ correction to the Berry phase term for the nematic order parameter comes from integrating out the spin-$3$ degrees of freedom.
There may exist a nonlinear theory that includes all of the higher spin fields and naturally provides the canonical structure on all shape deformations of the composite Fermi surface. Such theory may provide an understanding of the role of the $W_\infty$ algebra\cite{Cappelli1993infinite}, including the exact GMP algebra. The construction of such a theory is presently an exciting open problem.

\acknowledgements

We thank Steve Kivelson and Paul Wiegmann for useful discussions, and to Yizhi You for comments on an earlier version of the manuscript.
This work is supported, in part, by DOE grant No.\ DE-FG02-13ER41958, a Simons Investigator grant from the Simons Foundation, and by the University of Chicago Materials Research Science and Engineering Center, which is funded by the National Science Foundation under award No.\ DMR-1420709. D. X. N. is also supported, in part, by EPSRC grant EP/N01930X/1. AG was supported by the Leo Kadanoff Fellowship.

\appendix

\section{Dirac brackets}
\label{sec:DB}
Consider a classical theory where we can separate two sets of degrees of freedom,
$\xi_a$ and $\xi_A$, with the Hamitonian
\begin{equation}
  H = H(\xi_a, \xi_A).
\end{equation}
We assume that $\xi_A$ are much heavier than $\xi_a$.  The
equation of motion for $\xi_A$ is
\begin{equation}
  \dot \xi_A = \{\xi_A,\, H \} = \{ \xi_A,\, \xi_a \} \frac{\d H}{\d \xi_a}
  + \{ \xi_A,\, \xi_B \} \frac{\d H}{\d \xi_B} \,.
  \end{equation}
At low energies (energies much smaller than the energy scales of the
$u_A$ modes), the time derivative in this equation can be neglected, giving
\begin{equation}
  \frac{\d H}{\d \xi_A} = - \{ \xi,\, \xi \}^{-1}_{AB} \{\xi_B,\, \xi_a \} \frac{\d H}{\d\xi_a} \, ,
\end{equation}
where $\{ \xi,\, \xi \}^{-1}_{AB}$ is the matrix inverse of $\{\xi_A,\, \xi_B\}$.
Now we can write the equation of motion for $\xi_a$:
\begin{multline}
  \dot \xi_a =  \{ \xi_a,\, \xi_b \} \frac{\d H}{\d \xi_b} 
  + \{ \xi_a,\, \xi_A \} \frac{\d H}{\d \xi_A} \\
  =  \{ \xi_a,\, \xi_b \} \frac{\d H}{\d \xi_b} 
  -  \{ \xi_a,\, \xi_A \} \{ \xi\,, \xi \}_{AB}^{-1} \{ \xi_B,\, \xi_b \}  \frac{\d H}{\d \xi_b} \,.
\end{multline}
This can be rewritten as
\begin{equation}
   \dot \xi_a =  \{ \xi_a,\, \xi_b \}_{\rm D} \frac{\d H}{\d \xi_b} \equiv \{ \xi_a,\, H \}_{\rm D},
\end{equation}
where the Dirac  bracket is defined as
\begin{equation}
  \{ \xi_a,\, \xi_b \}_{\rm D} = \{ \xi_a,\, \xi_b \} 
  - \{\xi_a,\, \xi_A \} \{ \xi,\, \xi \}_{AB}^{-1} \{ \xi_B,\, \xi_b \}.
\end{equation} 

\section{Evaluating the Dirac bracket}
\label{sec:evalDB}

The commutators between $u_n$ can be written in momentum space as,
\begin{equation}
  [u_n(\q),\, u_{-m}(\q')] = C_{nm}(\q) (2\pi)^2\delta(\q+\q') ,
\end{equation}
with $C_{nm}$ is a matrix whose nonzero elements are
\begin{subequations}
\begin{align}
  C_{nn} &= - \frac{2\pi}{k_F^2} n b , \\
  C_{n,n+1} = C_{n+1,n} &= - \frac{2\pi}{k_F} q_{\bar z} ,\\
  C_{n,n-1} = C_{n-1,n} &= - \frac{2\pi}{k_F} q_z .
\end{align}
\end{subequations}
The Dirac bracket between $u_{\pm2}$ is then
\begin{equation}
  [u_2(\q),\, u_{-2}(\q')]_{\rm D} = (C_{22} - C_{23} C^{-1}_{33} C_{32} )(2\pi)^2\delta(\q+\q'),
\end{equation}
where $C^{-1}_{MN}$ is the inverse of the matrix $C_{MN}$, $M,N=3,4,\ldots$. To compute $C^{-1}_{33}$ it is sufficient to solve the system of linear equations
\begin{subequations}\label{eq:C-1}
\begin{align}
  C_{3N}^{\phantom{1}} x_N^{\phantom{1}} & = 1,\\
  C_{MN}^{\phantom{1}} x_N^{\phantom{1}} &= 0, \qquad M \ge 4.
\end{align}
\end{subequations}
Then $C^{-1}_{33}=x_3$.  Without losing generality we take $q_x=q$, $q_y=0$. The solution to Eq.~(\ref{eq:C-1}) is 
\begin{equation}
   x_n = \frac{k_F}{2\pi} \frac2q \frac{(-1)^n J_n(\mq)}{J_2(\mq)}\,,
\end{equation}
and from that we find Eq.~(\ref{eq:u2Dirac})

\section{Conventions}
\label{sec:conventions}
Here we summarize the complex notations. We define
\be
z = x+ i y\,,\qquad \bar z = x-iy\,.
\ee
The derivatives are defined as
\be
\partial_z = \partial = \frac{1}{2}(\partial_x - i \partial_y)\,, \qquad \partial_{\bar z} = \bar \partial =  \frac{1}{2}(\partial_x + i \partial_y)\,.
\ee
The Laplace operator is given by $ \Delta = 4 \p \bar \p$.
Components of any one-form are defined as
\be
A_z = A = A_x - iA_y\,,\qquad A_{\bar z} = \bar A =  A_x + iA_y\,.
\ee
Similarly for higher rank tensors we have
\bea
g_{zz} &=& g_{11} - g_{22} - i(g_{12} + g_{21})\,,\qquad g_{\bar z \bar z} = \bar{g}_{zz}\,,\\
 g_{z \bar z} &=& g_{11} + g_{22} + i(g_{12} - g_{21})\,.
\eea
With these definitions non-zero components of the Levi-Civita symbol are
\be
\epsilon_{z\bar z} = - \epsilon_{\bar z z} = 2i\,.
\ee
Any contraction of indices requires extra factors of $1/2$
 \be
 u_i v_i = \frac{1}{2} (u \bar v + \bar u v)\,. 
 \ee
The cross-product takes form
\be
\epsilon_{ij} u_i v_j = \frac{i}{2} (\bar u v - u \bar v)\,.
\ee
The exception for the contractions is a contraction with derivative, since derivative already has extra factor of $1/2$ in its definition
\be
\partial_i v_i = \p \bar v + \bar \p v\,,\quad \epsilon_{ij} \partial_i v_j = i(\bar \p v - \p \bar v)\,.
\ee
Consider a counterclockwise rotation by an angle $\phi$
\be
R =  \begin{pmatrix}
   \cos\phi   & -\sin\phi \\
   \sin\phi       & \cos\phi
\end{pmatrix}.
\ee 
Under this rotation we have
\bea
&&z \rightarrow e^{+i\phi} z\,, \quad \partial \rightarrow e^{-i\phi} \partial\,, \quad A_z \rightarrow e^{-i\phi}A_z\,,\\
 &&\quad g_{zz} \rightarrow e^{-2i\phi} g_{zz}\,,\quad g_{z\bar z} \rightarrow g_{z\bar z}.
\eea
Finally, we define the complex momentum  as
\be
k_z = k = k_1 - i k_2\,,\quad k_{\bar z} = \bar k = k_1 + ik_2\,.
\ee
Then the rules of Fourier transform are
\bea
&&\mathcal F [\partial f] = \frac{i k }{2} f\,, \qquad \mathcal F [\bar \partial f] = \frac{i \bar k }{2} f\,, \\
&& \mathcal F [\Delta f] = 4 F [\p \bar \p f] = -|k|^2 f .
\eea
Consider a unimodular symmetric tensor
\be
g = \exp \begin{pmatrix}
   Q_2   & Q_1 \\
   Q_1       & -Q_2
\end{pmatrix},\quad Q = Q_1+ iQ_2,\quad Q \rightarrow e^{-2i\phi} Q .
\ee
We find its complex components
\bea\nonumber
&&g_{zz} = -2 i Q \frac{\sinh|Q|}{|Q|} \approx -2iQ\,,\quad g_{\bar z \bar z} = 2 i \bar Q \frac{\sinh|Q|}{|Q|}\approx2 i \bar Q\,,\\
&& g_{z\bar z} = 2 \cosh|Q|\approx 2 + Q\bar Q\,, \quad |Q| = \sqrt{Q \bar Q} \,.
\eea

Next we find the linearized spin connection
\be
\omega_i = - \frac{1}{2} \epsilon_{jk} \partial_j g_{ki}\,.
\ee
Then
\be
\omega_z = - \frac{i}{2} \bar \p g_{zz} = -\bar \p Q \,, \quad \omega_{\bar z} = -\p \bar Q\,,
\ee
The Ricci scalar is
\be
R = 2 \epsilon_{ij} \p_i \omega_j = 2i(\bar \p \omega - \p \bar \omega) = 2i (\p^2 \bar Q - \bar \p^2 Q) .
\ee
Also
\be
\omega_0 = \frac{1}{2} \epsilon_{ij} g_{ik} \dot g_{kj} = \frac{i}{4} (\bar Q \dot Q - \dot{\bar Q} Q) .
\ee
The kinematic part of the action is
\bea\nonumber
&&S = \frac{\nu \varsigma}{2\pi}\! \int\! A d \hat \omega \\ \nonumber
&&=  \frac{i}{2} \frac{\nu\varsigma}{2\pi \ell^2}\!\int\! \bar Q \dot Q + i\frac{\nu \varsigma}{2\pi} \!\int\! A_0 (\partial^2 \bar Q - \bar \partial^2 Q) \\
&&+ \frac{i \nu \varsigma}{4\pi} \!\int\! A \p \bar Q - \bar A \bar \p Q.
\eea
The density is
\be
\rho = \frac{\nu \varsigma}{4\pi} \hat R =i \frac{\nu \varsigma}{2\pi} (\partial^2 \bar Q -\bar \partial^2 Q).
\ee

\section{Detailed derivation of the single-mode action (\ref{eq:LagSMA1})}
\label{sec:deriSMA}

In this Appendix, we will derive the equations \eqref{eq:LagSMA1} in detail. Combining Eqs.~\eqref{eq:rec1} and \eqref{eq:u2fin}, we obtain
\begin{equation}
\label{eq:1}
\left[\omega\pm f_2\omega_c G(\mq)\right]u_{\pm 2}=\mp \mq_{\pm}\omega \alpha_{\pm}, 
\end{equation}
where
\begin{equation}
G(\mq)=F(\mq)-\frac{\mq^2}{4}=\frac{\mq^2}{4}\frac{J_0(\mq)}{J_2(\mq)},
\end{equation} 
and we define 
\begin{equation}
\alpha_+=\delta \tilde{a}_++\frac{i\tilde E_+}{(2N{+}1)\omega_c}, \quad \alpha_{-}=\delta \tilde{a}_{-}-\frac{i\tilde E_{-}}{(2N{+}1)\omega_c}.
\end{equation}
 We can rewrite Eq.~\eqref{eq:u2fin} as 
\begin{equation}
\label{eq:2}
i\frac{\tilde E_{\pm}}{2N+1}=\mq_{\mp}f_2\omega_cu_{\pm 2}-\omega\alpha_{\pm}.
\end{equation}
We propose a quadratic Lagrangian which satisfies the field equations \eqref{eq:1} and \eqref{eq:2}
\begin{multline}\label{eq:3}
\cL= H(\mq)\Bigl[\omega u_2  u_{-2}-f_2\omega_c G(\mq)u_2u_{-2}
\\ -\ell(2N{+}1)\left( \omega q_-u_2  \alpha_{-}+\omega q_+u_{-2} \alpha_+\right) +K(\mq)\omega\alpha_+  \alpha_{-}\Bigr] \\
+\frac{i}{2\pi}\bigl(\alpha_z \tilde E_{\bar{z}}-\alpha_{\bar{z}}\tilde E_z\bigr)+\frac{i}{8\pi}\epsilon^{\mu\nu\rho}\cA_\mu q_\nu \cA_\rho-\frac{\bar b}{4\pi}A_0,
\end{multline} 
where $H(\mq)$ and $K(\mq)$ are 2 functions will be determined momentary. The last 3 terms of effective action \eqref{eq:3} come from the second line of equation \eqref{eq:Dirac2}  \footnote{The $\tilde E_i \tilde E^i$ terms is canceled by the shift of $\delta \tilde{a}_{\pm}$.}. The equations of motion (eom) for $u_{\pm 2}$ are the same as \eqref{eq:1}. The eom for $\alpha_z$ and $\alpha_{\bar{z}}$ are
\begin{equation}
\label{eq:4}
H(\mq)\left[\mp K(\mq)\omega \alpha_{\pm}-\omega\mq_{\mp} u_{\pm 2}\right]=\mp \frac{i}{2\pi}\tilde E_{\pm}.
\end{equation}
Combining Eqs.~\eqref{eq:1} and \eqref{eq:4} give us 
\begin{equation}
\label{eq:5}
H(\mq)\left[\mp \left(K(\mq)-\frac{\mq^2}{4}\right)\omega \alpha_{\pm}\pm \mq_{\mp}f_2 G(\mq)u_{\pm 2}\right]=\mp \frac{i}{2\pi}\tilde E_{\pm}.
\end{equation}
Comparing Eqs.~\eqref{eq:5} and \eqref{eq:2}, we obtain 
\begin{align}
\label{eq:6}
H(\mq)=\frac{2N+1}{2\pi G(\mq)}, \qquad K(\mq)=G(\mq)+\frac{\mq^2}{4}=F(\mq).
\end{align}
Substituting Eq.~\eqref{eq:6} into Eq.~\eqref{eq:3}, we arrive at the effective Lagrangian 
\begin{multline}\label{eq:Lag1}
\cL= \frac{2N{+}1}{2\pi G(\mq)} \Bigl[\omega u_2  u_{-2}-f_2\omega_c G(\mq)u_2u_{-2}
\\ .-\ell(2N{+}1)\left( \omega q_-u_2  \alpha_{-}+\omega q_+u_{-2} \alpha_+\right) +F(\mq)\omega\alpha_+  \alpha_{-}\Bigr] \\
+\frac{i}{2\pi}\bigl(\alpha_z \tilde E_{\bar{z}}-\alpha_{\bar{z}}\tilde E_z\bigr)+\frac{i}{8\pi}\epsilon^{\mu\nu\rho}\cA_\mu q_\nu \cA_\rho-\frac{\bar b}{4\pi}A_0.
\end{multline}  
 We can further integrate out $\alpha_{z}$ and $\alpha_{\bar{z}}$ by the equations of motion 
 \begin{align}
 \label{eq:az}
 \alpha_+&=-\frac{i}{2N+1}\frac{G(\mq)}{F(\mq)}\frac{\tilde E_+}{\omega}-\frac{\mq_{-}}{F(\mq)}u_2,\\
 \label{eq:azb}
 \alpha_{-} &=-\frac{i}{2N+1}\frac{G(\mq)}{F(\mq)}\frac{\tilde E_{-}}{\omega}+\frac{\mq_+}{F(\mq)}u_{-2},
 \end{align}
and obtain the action \eqref{eq:LagSMA1}.


\bibliography{BimetricCFL}

\end{document}